\documentstyle[prd,aps,eqsecnum]{revtex}

\begin{document}
\preprint{QMW-AU-96018 (March 1996)}
\draft

\title{\LARGE \bf Averaging Problem in General Relativity,\\
Macroscopic Gravity and\\
Using Einstein's Equations in Cosmology\thanks{This essay has received 
Honorable Mention from the Gravity Research Foundation for 1996}}

\author{Roustam M. Zalaletdinov\thanks{E-mail address: 
R.M.Zalaletdinov@qmw.ac.uk}\thanks{\'Sniadeccy Fellow (Cracow, Poland),
address from 10 May 1996 to 9 May 1997: N. Copernicus Astronomical Center, 
Polish Academy of Sciences, ul. Bartycka 18, 00-716 Warsaw, Poland; 
e-mail address: Roustam$_-$Zalaletdinov@camk.edu.pl}}

\address{School of Mathematical Sciences, Queen Mary \& Westfield College\\
University of London, Mile End Road, London E1 4NS, England, U.K.}

\address{Department of Theoretical Physics, Institute of Nuclear Physics\\
Uzbek Academy of Sciences, Tashkent 702132, Uzbekistan, C.I.S.}


\maketitle

\begin{abstract}
The averaging problem in general relativity is briefly discussed. A new
setting of the problem as that of macroscopic description of gravitation is
proposed. A covariant space-time averaging procedure is described. The
structure of the geometry of macroscopic space-time, which follows from
averaging Cartan's structure equations, is described and the correlation
tensors present in the theory are discussed. The macroscopic field equations
(averaged Einstein's equations) derived in the framework of the approach are
presented and their structure is analysed. The correspondence principle for
macroscopic gravity is formulated and a definition of the stress-energy
tensor for the macroscopic gravitational field is proposed. 
It is shown that the
physical meaning of using Einstein's equations with a hydrodynamic
stress-energy tensor in looking for cosmological models means neglecting all
gravitational field correlations. The system of macroscopic gravity
equations to be solved when the correlations are taken into consideration is
given and described.
\end{abstract}


\section{The averaging problem in general relativity}

The usual practice of using Einstein's equations in looking for cosmological
solutions is to assume that the real complex lumpy universe with a discrete
matter distribution (stars, galaxies, clusters of galaxies, etc.) can be
adequately approximated by a ``smoothed'', or hydrodynamic, stress-energy
tensor usually taken to be representable by a perfect fluid. 
Such an approximation
assumes an effective averaging of the discrete matter distribution. It is
tacitly assumed at the same time, and it is the essence of employing the
field equations of general relativity in modern cosmology, that
Einstein's equations remain unchanged in the structure of their field
operator under such an averaging. Apart from the question of whether or not
a hydrodynamic picture is satisfactory for the matter distribution in the
universe, such a procedure of employing Einstein's equations in cosmology
raises questions of principle which constitute the so-called averaging
problem in general relativity \cite{Shir-Fish:1962}-\cite{Zoto-Stoe:1992}
(see \cite{Elli:1984},\cite{Kras:1996} for a review and references). Indeed,
a correct statement of the problem requires the averaging out of Einstein
equations in both sides, in the matter source and in the field operator. And
it is the averaged equations which should then be solved to find the
relevant cosmological models\footnote{The averaging 
problem in general relativity applies also in cases of 
using Einstein's equations inside gravitating extended bodies where a 
hydrodynamic matter model is being used again. 
Similar arguments as in cosmology 
question here the applicability of Einstein's equations in such settings, but 
this is not discussed in this essay (see, however, \cite{Tava-Zala:1996}).}.
The results of averaging are expected to be far from trivial, since
Einstein's equations are highly non-linear, and the averaging is likely to
change their structure. Solving the averaged equations may therefore be
expected to bring about a new view on cosmology, which could in turn
alter our understanding and modify the predictions which are heavily based,
in modern cosmology, on the solutions of Einstein's equations.

This essay is aimed to give an overview of results on the averaging problem
within the macroscopic gravity approach proposed recently 
\cite{Zala:1992},\cite{Zala:1993}-\cite{Zala:pres}, and to reveal, 
within its context, the
physical meaning and the range of validity of using Einstein's equations
with an averaged, hydrodynamic stress-energy tensor in studies of
cosmological problems.

\section{A setting of the problem of macroscopic gravity}

The averaging problem in general relativity can be reformulated in a
broader context as the problem of macroscopic description of gravitation 
\cite{Zala:1992},\cite{Zala:1993}-\cite{Zala:1996}. The idea of macroscopic
gravity can be considered as an extension of Lorentz's idea, formulated
first for electrodynamics \cite{Lore:1916}, regarding the existence of two
levels, microscopic and macroscopic, of understanding classical physical
phenomena. The microscopic description deals with the matter structured by
discrete constituents, while the macroscopic description represents matter
from a hydrodynamic point of view. However, unlike electrodynamics where the
field operator is linear in the fields and it can be easily averaged out by
applying either space, time, or statistical averagings (or a combination of
them), and the remaining problem is the construction of models of continuous
electromagnetic media which relates to the structure of averaged
(macroscopic) current (see \cite{deGr-Sutt:1972} for discussion and
details), the problem of derivation of a macroscopic theory of gravity does
require one to overcome severe difficulties, even on the first step of
averaging the Einstein field operator. These are clearly connected with the
non-flat geometry underlying the general relativity theory and the nonlinear
character of the gravitational field, resulting in the 
need for considering the
field correlations. In such a setting Einstein's equations are to be
considered as the microscopic ones with a microscopic stress-energy tensor $
t_\beta ^{\alpha {\rm (micro)}}$, which may be considered as
well-grounded, for it is these equations that are believed to provide us with
an exact solution for the gravitational field of an isolated point mass
(Schwarzschild's solution)\footnote{There exists, 
however, another point of view where Einstein's equations
are considered to be macroscopic in their nature and they can be used 
consistently only with a continuous matter model. 
Such a setting requires one to pose a problem of finding corresponding 
equations for a microscopic theory of gravity, which would lead to 
Einstein's equations under an appropriate averaging procedure. 
This problem is similar to that which Lorentz had resolved by discovering
a microscopic theory of electromagnetism and showing Maxwell's 
theory to be its macroscopic version.}.

It should be pointed out also that macroscopic description has a direct
observational status since a (space-time) averaging procedure is a model of
a measurement procedure. Physically, a macroscopic theory has a direct
observational status and answers the questions of principle of which objects
can be observed in classical measurements of fields and matter (as induction
in electrodynamics, for example).

The overwhelming majority of approaches to the averaging problem \cite
{Kras:1996} has three main features: (I) they are {\em perturbative} (an
averaging of perturbed Einstein's equations is carried out); (II) they
follow the generally adopted way, evoked by Lorentz's approach to
electrodynamics, of {\em an averaging of Einstein's equations} to arrive at
the averaged field equations and to understand thereby a structure of
averaged gravity, and (III) no proposal has been made about the 
{\em correlation functions} which should 
inevitably emerge in an averaging of a
non-linear theory. The results derived failed to give and, can not in fact
give, a satisfactory solution to the problem since (I$^{\prime }$) any
perturbation analysis can not give information about an averaged geometry; 
(II$^{\prime }$) an averaging of Einstein's equations themselves can be easily
demonstrated to be insufficient because (III$^{\prime }$) they themselves
can not be used as a definition of the correlation functions if one wishes
to have them as {\em the field equations}. Indeed, consider the Einstein
equations in the mixed form 
\begin{equation}
\label{EE} g^{\alpha \epsilon }r_{\epsilon \beta }-\frac 12\delta _\beta
^\alpha g^{\mu \nu }r_{\mu \nu }=-\kappa t_\beta ^{\alpha {\rm (micro)}}\ , 
\end{equation}
which looks preferable for one must deal with only products of metric by
curvature, as otherwise, in their other forms, 
one faces triple products of
metric by metric by curvature. Suppose now that a space-time averaging
procedure $\langle \cdot \rangle $ for tensors on space-time has been
defined. Then the averaging of (\ref{EE}) brings 
\begin{equation}
\label{averEE:1} \langle g^{\alpha \epsilon }r_{\epsilon \beta }\rangle
-\frac 12\delta _\beta ^\alpha \langle g^{\mu \nu }r_{\mu \nu }\rangle
=-\kappa \langle t_\beta ^{\alpha {\rm (micro)}}\rangle \ , 
\end{equation}
which can be rewritten as 
\begin{equation}
\label{averEE:2} \langle g^{\alpha \epsilon }\rangle \langle r_{\epsilon
\beta }\rangle -\frac 12\delta _\beta ^\alpha \langle g^{\mu \nu }\rangle
\langle r_{\mu \nu }\rangle +C^{\alpha}_{\beta }=-\kappa \langle t_\beta
^{\alpha {\rm (micro)}}\rangle \ , 
\end{equation}
where $\langle g^{\alpha \beta }\rangle $ and $\langle r_{\alpha \beta
}\rangle $ denote the averaged inverse metric and the Ricci tensors, and $
C^{\alpha}_{\beta }$ stands for a correlation function, which is just the
difference between (\ref{averEE:1}) and (\ref{averEE:2}). The averaged
Einstein equations (\ref{averEE:2}) become now a definition of the
correlation function. In order to bring back their status of the field
equations one should define the object and find the properties and equations
for $C^{\alpha}_{\beta}$ by using some information outside the Einstein
equations (\ref{EE}).

To resolve all these problems it has been proposed in the macroscopic gravity
approach that in order to derive the form of the averaged Einstein operator
one should study, first of all, the problem of how {\em to average out a
(pseudo-)Riemannian space-time }itself, i.e. Cartan's structure equations 
\cite{Cart:1928},\cite{Koba-Nomi:1963} describing the structure of a
(pseudo)-Riemannian geometry. While doing this it is necessary to understand
which averaged geometrical object - metric, connection, or curvature - can
characterize an averaged space-time. Another necessary part of such an
approach is the splitting of the averages of products of the objects, being
found in averaging out Cartan's equations. This is the problem of {\em 
introducing the correlation functions}. Upon deriving the structure
equations for the averaged manifold, the Einstein equations which are known
to be additional conditions to Cartan's equations, can successfully be
averaged out. Such an approach to formulate a macroscopic theory of gravity
is essentially {\em non-perturbative} and provides us with both {\em the
geometry }underlying the macroscopic gravitational phenomena and the {\em 
macroscopic (averaged) field equations}.

\section{A space-time averaging procedure}

Let us remind the definition of the space-time averages adopted in
macroscopic gravity \cite{Zala:1992},\cite{Zala:1993}. The space-time
averaging procedure is a generalization of the space-time averaging
procedure used in electrodynamics (see, for example, 
\cite{Nova:1955}-\cite{Inga-Jami:1985})\footnote{Averaging 
procedures applied in the derivation of macroscopic 
electrodynamics have incorporated
one or more combinations of three types of averaging procedures, 
viz., spatial, time, ensemble averagings. The discussion  
\cite{deGr:1969}-\cite{Jack:1975} seems to lead to a conclusion 
that a space averaging is always necessary and unavoidable in all macroscopic 
settings. Further application of either a time or statistical averaging 
depends on the problem of interest, or even could be unnecessary if there is 
no time periodicity in the problem under study. In generalization to 
a space-time formulation of macroscopic theories it is reasonable 
and mathematically preferable, in our opinion, 
to consider a space-time averaging procedure as fundamental one, with 
spatial averages taken as a projection of the space-time averages on 
hypersurfaces if the microscopic physics possesses no regularities 
and periodicity along a time-like direction. 
It is also known that it is space-time averages of physical fields that have 
the physical meaning \cite{Bohr-Rose:1933},\cite{DeWi:1962}. An alternative 
approach to that adopted here may be development of a space-time ensemble 
averaging procedure, though space-time procedures have its own advantages.}
and it is based on the concept of Lie-dragging of averaging 
regions\footnote{A definition of Lie-dragging (or, dragging) of a 
manifold's region of along a vector field (congruence), that is a 
mapping of the region into itself along the vector field (congruence), 
can be found in any standard textbook on differential geometry 
(see, for example, \cite{Koba-Nomi:1963},\cite{Schu:1980}). 
Hereafter Lie-draggings of regions 
are supposed to be diffeomorphisms of the same differentiability class as 
that of the space-time manifold 
$({\cal M}$,$g_{\alpha \beta})$.}, 
which makes it valid for any differentiable
manifolds with a volume $n$-form. 
Chosen a compact region $\Sigma \subset {\cal M}$ of a
differentiable space-time manifold $({\cal M}$,$g_{\alpha \beta })$ and a
supporting point $x\in \Sigma $ to which the average value will be
prescribed, the average value of an object (tensor, geometric objects, etc.) 
$p_\beta ^\alpha (x),\,x\in {\cal M\ }$, over a region $\Sigma $ at the
supporting point $x\in \Sigma $ is defined as 
\begin{equation}
\label{defaver} \bar p_\beta ^\alpha (x)=\frac 1{V_\Sigma }\int_\Sigma {\bf p
} _\beta ^\alpha (x,x^{\prime })\sqrt{-g^{\prime }}d^4x^{\prime }\equiv
\langle {\bf p}_\beta ^\alpha \rangle \ , 
\end{equation}
where $V_\Sigma $ is the 4-volume of the region $\Sigma $, 
\begin{equation}
\label{volume} V_\Sigma =\int_\Sigma \sqrt{-g}d^4x \ , 
\end{equation}
with the averaged object $\bar p_\beta ^\alpha $ keeping the same tensorial
character as $p_\beta ^\alpha $. Here the integration is carried out over
all points $x^{\prime }\in \Sigma $ , $g^{\prime }=\det (g_{\alpha \beta
}(x^{\prime }))$, and the boldface object ${\bf p}_\beta ^\alpha
(x,x^{\prime })$ in the integrand of (\ref{defaver}) is a bilocal extension
of the object $p_\beta ^\alpha (x)$, ${\bf p}_\beta ^\alpha (x,x^{\prime })= 
{\cal A}_{\mu ^{\prime }}^\alpha (x,x^{\prime })p_{\nu ^{\prime }}^{\mu
^{\prime }}(x^{\prime }){\cal A}_\beta ^{\nu ^{\prime }}(x^{\prime },x)$, by
means of bilocal averaging operators ${\cal A}_{\beta ^{\prime }}^\alpha
(x,x^{\prime })$ and ${\cal A}_\beta ^{\alpha ^{\prime }}(x^{\prime },x)$.
The averaging scheme is covariant and linear by its structure with
corresponding algebraic properties. The operator ${\cal A}_{\beta ^{\prime
}}^\alpha $ is supposed to exist on ${\cal U}\subset {\cal M}$, $x,x^{\prime
}\in {\cal U}$ , and to have the following properties\footnote{They are 
a formalization of the properties of the
space-time averages in macroscopic electrodynamics in the language of 
bilocal kernels.}: ({\em i}) idempotency, ${\cal A}_{\beta ^{\prime
}}^\alpha {\cal A}_{\gamma ^{\prime \prime }}^{\beta ^{\prime }}={\cal A}
_{\gamma ^{\prime \prime }}^\alpha $, that results in the idempotency of the
averages, $\overline{\overline {p}}_\beta ^\alpha =\bar p_\beta ^\alpha $; 
({\em ii}) the coincidence 
limit $\lim _{x^{\prime }\rightarrow x}{\cal A}_{\beta
^{\prime }}^\alpha (x,x^{\prime })=\delta _\beta ^\alpha $, with both
properties defining ${\cal A}_\beta ^{\alpha ^{\prime }}$ as an inverse
operator to ${\cal A}_{\beta ^{\prime }}^\alpha $, ${\cal A}_{\beta ^{\prime
}}^\alpha {\cal A}_\gamma ^{\beta ^{\prime }}=\delta _\gamma ^\alpha $ and $
{\cal A}_{\beta ^{\prime }}^\alpha {\cal A}_\alpha ^{\gamma ^{\prime
}}=\delta _{\beta ^{\prime }}^{\gamma ^{\prime }}$. By assigning an
averaging region $\Sigma _x$ near each point $x$ of ${\cal M}$ and by
defining a law of the correspondence between the neighbouring averaging
regions as Lie-dragging of regions by means of another bilocal operator 
${\cal W}_\beta ^{\alpha ^{\prime }}(x^{\prime },x)$ , one can define
directional, partial and covariant derivatives of average fields.
Requirements that ({\em a}) all averaging regions have the same value 
of volume (\ref{volume}), $V_{\Sigma _x}={\rm const}$, 
for all $x\in {\cal M}$ and 
({\em b}) a region remains the same after its Lie-dragging along a circuit 
constructed from vectors with vanishing Lie brackets, bring about special 
conditions on the two bivectors, taken for simplicity as 
${\cal W}_\beta ^{\alpha ^{\prime}}={\cal A}_\beta ^{\alpha ^{\prime }}$, 
\begin{equation}
\label{diffW:2} {\cal W}_{\beta ; \alpha ^{\prime }}^{\alpha ^{\prime }}=0\ ,
\end{equation}
where the semicolon denotes a covariant derivative with respect to 
connection coefficients $\gamma^\alpha_{\beta\delta}$, and
\begin{equation}
\label{diffW:1} {\cal W}_{[\beta ,\gamma ]}^{\alpha ^{\prime }}+
{\cal W}_{[ \beta ,\delta ^{\prime }}^{\alpha ^{\prime }}
{\cal W}_{\gamma ]}^{\delta^{\prime }}=0\, \ .
\end{equation}
The first requirement (\ref{diffW:2}) means that
the averaging regions' volume is a free parameter of the procedure and this
assumes physically that measurements carried out on a differentiable
manifold by an observer with a given measurement system can be characterized
by an invariant number $V_{\Sigma _x}={\rm const}$ which is to be fixed, or
chosen, for a situation under consideration. The second requirement 
(\ref{diffW:1}) means that the covering of a manifold by averaging 
regions is defined uniquely,
which results in a non-trivial analytical property of the averages 
(\ref{defaver}) as single-valued {\em local } functions of the 
supporting point, $\overline{p}_{\beta ,[\mu \nu ]}^\alpha =0$ and the
standard calculus is therefore applicable to deal with them. These
requirements are particular differential conditions on the
bivectors, which means geometrically that the bivectors act on ${\cal M}$ as
volume-preserving diffeomorphisms holonomic in a defined bilocal sense
(biholonomic, see \cite{Zala:1993}). The bivectors satisfying 
(\ref{diffW:2}), (\ref{diffW:1}) together with the algebraic properties
({\em i}) and ({\em ii}) above have been shown 
\cite{Zala:1992},\cite{Zala:1993} 
to exist on an arbitrary differentiable manifold with a volume $n$-form 
and solutions
of the system of algebraic and partial differential equations have been 
found for the bivectors in a factorized form. These solutions are 
bilocal products of a vector basis $e_i ^{\alpha}$ 
with constant anholonomicity coefficients, taken at one point, 
by its dual basis $e_\mu ^k$ in another point,
${\cal W}_\beta ^{\alpha ^{\prime }}(x^{\prime },x) =
e_i ^{\alpha ^{\prime }}(x^{\prime }) e_\beta ^i(x)$.
It should be noted here that solutions of such structure are in fact 
the general solution for equation (\ref{diffW:1}), for a bilocal 
operator satisfying the algebraic properties ({\em i}) and ({\em ii})
can be shown to be idempotent iff it is factorized into the bilocal product
\cite{Zala:Unpub}. 
In the simplest case when $e_i ^{\alpha}$ is a coordinate basis, 
$e_i ^{\alpha}=\partial x^{\alpha} / {\partial \phi^i}$ 
where $\phi^i(x)$ are four arbitrary scalar functions, and the bivector
takes the form
\begin{equation}
\label{phiW} {\cal W}_\beta ^{\alpha ^{\prime }}(x^{\prime },x) =
\frac {\partial x^{\alpha ^\prime}} {\partial \phi^i}
\frac {\partial \phi^i} {\partial x^ \beta} \ ,
\end{equation}
a proper coordinate system formed by $\phi^i$'s is a 
volume-preserving coordinate system, in which 
${\cal W}_\beta ^{\alpha ^{\prime}}(x^{\prime },x)=
\delta _\beta ^{\alpha ^{\prime }}$ and $\det (g_{\alpha\beta })={\rm const}$ 
and all definitions and relations within the averaging scheme acquire 
especially simple form. This coordinate system is an analogue of Cartesian 
coordinates in the flat space-time case. 

The basic commutation formula for the averaging 
and exterior derivative, obtained in the framework of the 
formalism, has the following form \cite{Zala:1993}:
\begin{equation}
\label{commut} {\rm d}\bar p_\beta ^\alpha (x)=\langle {\rm d\negthinspace
\negthinspace \negthinspace ^{-}}{\bf p}_\beta ^\alpha \rangle \, , 
\end{equation}
where ${\rm d\negthinspace \negthinspace \negthinspace ^{-}}$ is a bilocal
exterior derivative.

It should be pointed out here that such pretendents to be averaging 
kernels as either the parallel transportation bivector, or the one 
constructed by taking derivatives of Synge's world function in two 
different points 
\cite{DeWi:1962},\cite{Syng:1960} {\em do not satisfy} the conditions 
(\ref{diffW:2}) and (\ref{diffW:1}), with the latter not satisfying also 
the idempotency condition ({\em i}). Those objects are well-defined as
single-valued functions of their arguments only
for Whitehead's normal convex neighbourhoods 
(where there exists one and only one
geodesic between each pair of points), which makes problematic their
use for averaging the gravitational field over the space-time
regions with matter. Such cases are of the most interest from the physical 
point of view and a possibility to be able to treat them is an 
ultimate goal in the framework of the macroscopic description of classical
fields. Just to make use of such bivectors within the formalism makes many 
things very complicated \cite{Zala:Unpub}
(for example, the commutation formula (\ref{commut}) loses 
its transparent meaning and simple form). If, nevertheless, one decides to
utilize the parallel 
transportation bivector, ${\cal W}_\beta ^{\alpha ^{\prime
}}(x^{\prime },x)=g_\beta ^{\alpha ^{\prime
}}(x^{\prime },x)$, and to be restricted to the normal convex 
neighbourhoods, it is necessary to require the parallelly transported bases 
to have the constant anholonomicity coefficients to fulfill the 
condition (\ref{diffW:1}). The additional requirement for the operator 
$g_\beta ^{\alpha ^{\prime}}(x^{\prime },x)$ to be 
volume-preserving (\ref{diffW:2}) leads \cite{Zala:Unpub} 
to the class of D'Atri spaces (see, for example, \cite{Tric-Vanh:1983}
for definitions and relevant results), 
which is a special class of (pseudo)-Riemannian spaces with 
particular restrictions of their curvature. On the contrary, 
the volume-preserving bases 
$e_i ^{\alpha}(x)$ with constant anholonomicity coefficients defining
the averaging and coordination operators in the space-time averaging scheme
adopted in macroscopic gravity, exist always, at least, locally
\footnote{The space-time averaging scheme being considered here is
essentially of local character in the sense that the average values are
defined by (\ref{defaver}) over local regions $\Sigma$ of a microscopic
manifold $\cal M$, and thereby the average fields are defined locally on 
$\cal U$ with the topological and differentiable structure of $\cal M$ 
remaining unchanged. Such local character of the macroscopic picture
is dictated, first of all, from the physical point of view, by our 
experience  and observations which show that physical quantities are 
represented by local functions determined by means of measurements which
are themselves fundamentally of local character (i.e., a measurement 
of a physical quantity is
carried out always during a {\em finite} time period over a {\em finite} space
region, to be {\em small} compared with the characteristic extension 
of the system under interest and its time of existence). Thus, from 
the mathematical point of view, to describe such objects adequately, 
it is sufficient to
formulate a calculus of the averages on a differentiable manifold. If such
a calculus is formulated, a definition of an average field globally can be
done in the same way as one constructs a global field on a manifold if
it possesses a non-trivial topology.}, 
on any (pseudo)-Riemannian space-times without any restrictions on 
the curvature\footnote{It follows from the well-known fact that on any 
(pseudo)-Riemannian space there always exists a coordinate system 
in which the 
connection coefficients $\gamma^\alpha_{\beta \delta}$ have
$\gamma^\alpha_{\beta \alpha} = 0$, or, what is the same, 
$\det (g_{\alpha\beta })={\rm const}$.}. This averaging scheme, 
therefore, in addition
to the possibility to define averaged fields (\ref{defaver}) with 
reasonable algebraic and differential properties, is applicable (locally, 
as discussed above) on any space-time manifold. This is an essential
advantage of the scheme, which allows one to consider it, as well as the 
results of its application for the space-time averaging of 
(pseudo)-Riemannian geometry and general relativity, as being generic
from both geometrical and physical points 
of view\footnote{This analysis gives an indication, in our opinion, that 
using such bivectors as, for instance, the parallel transportation 
bivector, for the averaging of 
general relativity implies another set of
basic assumptions about the nature of averaged gravity and the character of 
space-time measurements \cite{Zala:Prog}. It also may be related 
to a quantum regime of gravitation as a physical setting 
where such averaging operators are more adequate (see, for example, 
\cite{DeWi:1962}). For formulation of a classical macroscopic theory 
of gravity, the proposed space-time averaging procedure based 
on a Lie-dragging model of 
space-time measurements is relevant and it is a simplest generalization
of the flat space-time procedure adopted in hydrodynamics and 
macroscopic electrodynamics.}.

The averages (\ref{defaver}) of products and corresponding correlation
functions are taken to be one-point functions of supporting point $x$, which
means that the approach developed is related to the equilibrium macroscopic
gravitational processes (for a non-equilibrium theory \cite{Zala:Prog} it is
necessary to introduce many-points version of averaged products and
correlation functions).

\section{The geometry of macroscopic space-time}
\label{geometry}

Resulting from the averaging out of Cartan's structure equations, the
geometry of the averaged (macroscopic) space-time has the following
structure \cite{Zala:1992},\cite{Zala:1993}. The average $\overline{{\cal F}}
^\alpha {}_{\beta \gamma }=\langle {\cal F}^\alpha {}_{\beta \gamma }\rangle 
$ of the microscopic Levi-Civita connection\footnote{Here 
${\cal F}^\alpha{}_{\beta\gamma}$ is a bilocal extension of the connection
coefficients $\gamma^\alpha{}_{\beta\gamma}$ \cite{Zala:1992}.} $\gamma
^\alpha {}_{\beta \gamma }$, is supposed to be Levi-Civita's connection of
the averaged space-time. A metric tensor $G_{\alpha \beta }$ always exists
locally due to Frobenius' theorem with given $\overline{{\cal F}}^\alpha
{}_{\beta \gamma }$ \cite{Koba-Nomi:1963}, and $G_{\alpha \beta }$ is
considered to be the macroscopic metric tensor. There are two curvature
tensors, $M^\alpha {}_{\beta \gamma \delta }$ and $R^\alpha {}_{\beta \gamma
\delta }$, Riemannian and non-Riemannian, respectively. The Riemannian
curvature tensor $M^\alpha {}_{\beta \gamma \delta }$ corresponds to the
Levi-Civita connection $\overline{{\cal F}}^\alpha {}_{\beta \gamma }$.
The second curvature is assumed to correspond to 
the average curvature tensor $
\bar r^\alpha {}_{\beta \gamma \delta }=R^\alpha {}_{\beta \gamma \delta }$
for another symmetric connection $\Pi ^\alpha {}_{\beta \gamma }$ which 
is non-metric (i.e. the connection is incompatible with the metric tensor
$G_{\alpha\beta}$), and
the average curvature is non-Riemannian in that sense. There is
a remarkable relation between the two curvature tensors, 
which results from averaging
the second Cartan equation\footnote{Note that underlined 
indices are not affected by  
antisymmetization.} $r^\alpha {}_{\beta \gamma \delta }=2\gamma ^\alpha
{}_{\beta [\delta ,\gamma ]}+2\gamma ^\alpha {}_{\epsilon [\gamma }\gamma _{
\underline{\beta }\delta ]}^\epsilon $ 
\begin{equation}
\label{structure}R^\alpha {}_{\beta \rho \sigma }=M^\alpha {}_{\beta \rho
\sigma }+Q^\alpha {}_{\beta \rho \sigma }\ .
\end{equation}

This relation is of the form of a constitutive relation between the
induction, $M^\alpha {}_{\beta \rho \sigma }$, and average field, $R^\alpha
{}_{\beta \rho \sigma }$, with $Q^\alpha {}_{\beta \rho \sigma }$ standing
for the polarization tensor (defined below in Eq. (\ref{eq:Z})). The origin
of this {\it geometric relation} lies in the simple, but non-trivial,
geometric fact of the non-linear definition of the affine curvature in terms
of connection, which results in the curvature determined by the average
connection not being equal to the average curvature. The curvature tensors
satisfy the corresponding Bianchi identities, the Bianchi identities for $
M^\alpha {}_{\beta \rho \sigma }$ resulting from averaging out the
microscopic ones \cite{Zala:1992},\cite{Zala:1993}.

There is an affine deformation tensor $A^\alpha {}_{\beta \gamma }=\overline{
{\cal F}}^\alpha {}_{\beta \gamma }-\Pi ^\alpha {}_{\beta \gamma }$ in this
geometry, which plays the role of the polarization potential. It satisfies
the partial differential 
equation \cite{Zala:1992},\cite{Zala:1993}\footnote{The covariant 
derivatives with respect to the
connections $\Pi ^\alpha {}_{\beta \gamma }$ and $\overline{{\cal F}}^\alpha
{}_{\beta \gamma }$ are denoted as $|$ and $\parallel $, respectively.} 
\begin{equation}
\label{eq:A} A^\alpha {}_{\beta [\sigma \parallel \rho ]}-A^\alpha
{}_{\epsilon [\rho }A^\epsilon {}_{\underline{\beta }\sigma ]}=-\frac
12Q^\alpha {}_{\beta \rho \sigma } \ , 
\end{equation}
which is always integrable with necessity on an arbitrary averaged manifold.
The tensor $A^\alpha {}_{\beta \gamma }$ therefore exists and the theory is
not empty.

In addition to these objects, in a 4-dimensional space-time there are three
correlation tensors. The correlation 2-form $Z^\alpha {}_{\beta [\gamma
}{}^\mu {}_{\underline{\nu }\sigma ]}(x)$ is defined as 
\begin{equation}
\label{eq:Z}Z^\alpha {}_{\beta [\gamma }{}^\mu {}_{\underline{\nu }\sigma
]}=\langle {\cal F}^\alpha {}_{\beta [\gamma }
{\cal F}^\mu {}_{\underline{\nu }\sigma ]}\rangle-
\overline{{\cal F}}^\alpha {}_{\beta [\gamma }
\overline{{\cal F}}^\mu {}_{\underline{\nu }\sigma ]}\ ,
\end{equation}
with $Q^\alpha {}_{\beta \gamma \lambda }=-2Z^\delta {}_{\beta [\gamma
}{}^\alpha {}_{\underline{\delta }\lambda ]}$ (note the change in the sign
in the definition of $Q^\alpha {}_{\beta \gamma \lambda }$ here and in \cite
{Zala:1995}-\cite{Zala:pres} as compared with \cite{Zala:1992},\cite
{Zala:1993},\cite{Zala:1994}), and the correlation 3-form $Y^\alpha
{}_{\beta [\gamma }{}^\mu {}_{\underline{\nu }\sigma }{}^\theta {}_{
\underline{\kappa }\pi ]}(x)$ and 4-form $X^\alpha {}_{\beta [\gamma }{}^\mu
{}_{\underline{\nu }\sigma }{}^\theta {}_{\underline{\kappa }\pi }{}^\tau
{}_{\underline{\varphi }\psi ]}(x)$ which are defined analogously 
to (\ref{eq:Z}) as triple and quadruple connection correlation tensors. 
These {\it tensors} are
constructed from the connection coefficients and the number of them is {\em 
finite} since the dimension of space-time is finite.

In their geometrical sense the correlation functions are non-trivial
generalizations of the concept of the affine curvature tensor. The geometric
meaning of the trace part $Q^\alpha {}_{\beta \gamma \lambda }$ of $Z^\alpha
{}_{\beta [\gamma }{}^\mu {}_{\underline{\nu }\sigma ]}$ can be shown to be
the difference between the defect $\delta v^\alpha $ for a vector $v^\alpha $
after its parallel transportation along a circuit with square $\Delta \sigma
^{\mu \nu }$ in a microscopic manifold with the curvature $r^\alpha
{}_{\beta \gamma \delta }$ of the connection $\gamma ^\alpha {}_{\beta
\gamma }$ with its subsequent averaging, and the defect $\delta \bar
v^\alpha $ for the averaged vector $\bar v^\alpha $ after its parallel
transportation along the same circuit in a macroscopic manifold with the
induction curvature $M^\alpha {}_{\beta \gamma \delta }$ of the averaged
connection $\overline{{\cal F}}^\alpha {}_{\beta \gamma }$ 
\begin{equation}
\label{defect} \langle \delta {\bf v}^\alpha \rangle -\delta \bar v^\alpha
=Q^\alpha {}_{\beta \mu \nu }\bar v^\beta \Delta \sigma ^{\mu \nu } \ . 
\end{equation}
The geometrical meaning of the tensor $Z^\alpha {}_{\beta [\gamma }{}^\mu
{}_{\underline{\nu }\sigma ]}$ itself and the tensors $Y^\alpha {}_{\beta
[\gamma }{}^\mu {}_{\underline{\nu }\sigma }{}^\theta {}_{\underline{\kappa }
\pi ]}$ and $X^\alpha {}_{\beta [\gamma }{}^\mu {}_{\underline{\nu }\sigma
}{}^\theta {}_{\underline{\kappa }\pi }{}^\tau {}_{\underline{\varphi }\psi
]}$ can be understood as generalization of (\ref{defect}) in the fibre
bundle picture \cite{Zala:Prog} of the space-time averaging scheme. There
are the structure equations for the correlation tensors. In the simplest
case when the other correlations tensors are put to zero, 
the structure equation
for the tensor $Z^\alpha {}_{\beta [\gamma }{}^\mu {}_{\underline{\nu }
\sigma ]}$ reads 
\begin{equation}
\label{eq:strZ} Z^\alpha {}_{\beta [\gamma }{}^\mu {}_{\underline{\nu }
\sigma \parallel \lambda ]}=0\ , 
\end{equation}
which is an analogue of the Bianchi identities, 
with the corresponding integrability
conditions \cite{Zala:1992},\cite{Zala:1993}.

The macroscopic metric tensor $G_{\alpha \beta }$ of the averaged space-time
is compatible with the Levi-Civita connection $\overline{{\cal F}}^\alpha
{}_{\beta \gamma }$, $G_{\alpha \beta \parallel \gamma }=0$. At the same
time the metric $G_{\alpha \beta }$ is incompatible with the connection $\Pi
^\alpha {}_{\beta \gamma }$, $G_{\alpha \beta |\gamma }=N_{\alpha \beta
\gamma }$ where $N_{\alpha \beta \gamma }$ is the non-metricity object. The
averaged metric tensors $\bar g_{\alpha \beta }\ne G_{\alpha \beta }$ and $
\bar g^{\alpha \beta }\ne G^{\alpha \beta }$ in general and they are not
metric tensors any more, i.e. $\bar g^{\alpha \beta }\bar g_{\beta \gamma
}\ne \delta _\gamma ^\alpha $, although the averaging has been shown to keep
them covariantly constant, $\bar g_{\alpha \beta \parallel \gamma }=0$ and ${
\bar g^{\alpha \beta }}{}_{\parallel \gamma }=0$, as a consequence of the
splitting rule 
\begin{equation}
\label{aver:gF}\langle {\bf c}_{\nu ...}^{\mu ...}{\cal F}^\alpha {}_{\beta
\gamma }\rangle =\bar c_{\nu ...}^{\mu ...}\overline{{\cal F}}^\alpha
{}_{\beta \gamma }\ ,
\end{equation}
which is assumed to hold for any covariantly constant tensors, Killing
vectors and tensors, and similar objects denoted as $\bar c_{\nu ...}^{\mu
...}$. It means geometrically that the averaging preserves the symmetries of
microscopic space-time. Due to the structure of the relations between $\bar
g_{\alpha \beta }$, $G_{\alpha \beta }$ and $M^\alpha {}_{\beta \gamma
\delta }$ one can always put, in addition, 
\begin{equation}
\label{eq:gbarG}\bar g_{\alpha \beta }=G_{\alpha \beta }\ .
\end{equation}
The covariantly constant symmetric tensor $\bar g^{\alpha \beta }$ is then
an object of the theory and it is convenient to define a tensor $U^{\alpha
\beta }=\bar g^{\alpha \beta }-G^{\alpha \beta }$ which is covariantly
constant, 
\begin{equation}
\label{eq:U}U{^{\alpha \beta }}_{\parallel \gamma }=0\ .
\end{equation}
The meaning of the tensor $U^{\alpha \beta }$ is that it describes the
algebraic metric correlations (all other metric correlations are contained
in the correlation tensors $Z^\alpha {}_{\beta [\gamma }{}^\mu {}_{
\underline{\nu }\sigma ]}$, $Y^\alpha {}_{\beta [\gamma }{}^\mu {}_{
\underline{\nu }\sigma }{}^\theta {}_{\underline{\kappa }\pi ]}$ and $
X^\alpha {}_{\beta [\gamma }{}^\mu {}_{\underline{\nu }\sigma }{}^\theta {}_{
\underline{\kappa }\pi }{}^\tau {}_{\underline{\varphi }\psi ]}$) in this
geometry\footnote{It should be noted that one can use both
tensors $\bar g_{\alpha \beta }$ and $\bar g^{\alpha \beta }$ without fixing
the ansatz (\ref{eq:gbarG}) and defining $U^{\alpha \beta }$, but it
simplifies the formalism without loss of generality and restriction of the
geometric content of the metric part of this geometry.}. The covariantly
constant correlation tensor $\Delta _\beta ^\alpha (x)$, defined by the
relation
\begin{equation}
\label{corr:metric}\delta _\beta ^\alpha =\langle {\bf g}^{\alpha \epsilon }
{\bf g}_{\epsilon \beta }\rangle =\bar g^{\alpha \epsilon }\bar g_{\epsilon
\beta }+\Delta _\beta ^\alpha\ ,
\end{equation}
can be shown to have the following structure when taking 
(\ref{eq:gbarG}) into account:
\begin{equation}
\label{corr:metric2}\Delta _\beta ^\alpha =-U^{\alpha \epsilon }G_{\epsilon
\beta }\ .
\end{equation}
The presence of the tensor $\bar g^{\alpha \beta }$, or $U^{\alpha \beta }$,
has been proved to make the macroscopic space-time reducible due to a
classification theorem \cite{Zala:1992},\cite{Zala:1993} of all possible
macroscopic space-times according to Petrov's types of the induction tensor
and kinds of the macroscopic metric tensor reducibility. This theorem
determines also the algebraic structure of the tensor $U^{\alpha \beta }$ in
terms of covariantly constant vectors and symmetric idempotent tensors.

The geometry of averaged space-time is a new geometry being a non-trivial
generalization of the metric affine connection one and it is this geometry
that underlies the macroscopic theory of gravity.

\section{The field equations of macroscopic gravity}

The splitting rule for the average of the product of 
{\it metric times curvature} derived in the theory 
\begin{equation}
\label{eq:gr}\langle {\bf r}^\alpha {}_{\beta \gamma \lambda }{\bf g}
^{\epsilon \rho }\rangle -R^\alpha {}_{\beta \gamma \lambda }\bar
g^{\epsilon \rho }=-2Z^\alpha {}_{\beta [\gamma }{}^\epsilon {}_{\underline{
\delta }\lambda ]}\bar g^{\delta \rho }-2Z^\alpha {}_{\beta [\gamma }{}^\rho
{}_{\underline{\delta }\lambda ]}\bar g^{\epsilon \delta }
\end{equation}
plays the most important role in macroscopic gravity, for it is {\it the
only rule }needed for the Einstein equations (\ref{EE})\ to be averaged out.
The result is the macroscopic field equations \cite{Zala:1992} 
\begin{equation}
\label{eq:averEE}\bar g^{\alpha \epsilon }M_{\epsilon \beta }-\frac 12\delta
_\beta ^\alpha \bar g^{\mu \nu }M_{\mu \nu }=-\kappa \langle {\bf t}_\beta
^{\alpha {\rm (micro)}}\rangle +(Z^\alpha {}_{\mu \nu \beta }-\frac 12\delta
_\beta ^\alpha Q_{\mu \nu })\bar g^{\mu \nu }
\end{equation}
which are not Riemannian in their geometrical meaning, though $\bar g^{\mu
\nu }$ is covariantly constant, $M_{\mu \nu }$ is the Ricci tensor of the
Riemannian curvature $M^\alpha {}_{\beta \gamma \delta }$, and the
divergence of the left-hand side of (\ref{eq:averEE}) vanishes \cite
{Zala:1992}, yielding the equations of motion, or the conservation law for
averaged matter {\em together} with correlation terms, 
\begin{equation}
\label{eq:motion}\kappa \langle {\bf t}_\beta ^{\alpha {\rm (micro)}}\rangle
_{\parallel \alpha }=(Z^\alpha {}_{\mu \nu \beta \parallel \alpha }-\frac
12Q_{\mu \nu \parallel \beta })\bar g^{\mu \nu }\ .
\end{equation}
Here $Z^\alpha {}_{\mu \nu \beta }=2Z^\alpha {}_{\mu [\epsilon }{}^\epsilon
{}_{\underline{\nu }\beta ]}$ is a Ricci-tensor like object for the
correlation tensor $Z^\alpha {}_{\beta [\gamma }{}^\mu {}_{\underline{\nu }
\lambda ]}$, $Q_{\mu \nu }=Q^\epsilon {}_{\mu \nu \epsilon }$, and $\langle 
{\bf t}_\beta ^{\alpha {\rm (micro)}}\rangle $ stands for an averaged
stress-energy tensor. The problem of calculation, or construction, of $
\langle {\bf t}_\beta ^{\alpha {\rm (micro)}}\rangle $ from a given
microscopic stress-energy tensor $t_\beta ^{\alpha {\rm (micro)}}$
constitutes the problem of construction of macroscopic gravitating media.
This is still an open problem in general relativity\footnote{Unlike 
electrodynamics 
where there is the whole industry of modelling
electromagnetic continuous media, though there are still a number of unsolved 
problems of principle regarding the derivation of material relations 
from the microscopic equations of motion (see, 
for example, \cite{deGr:1969}).} (see an attempt in \cite{Szek:1971} and
also relevant approaches \cite{Hava:1979},\cite{Dixo:1979}). The usual
practice in cosmology is to assume phenomenologically one or 
another form of smoothed,
hydrodynamic stress-energy tensors on the basis of observational data, and
the results of theoretical analysis are then compared with the data again to
conclude about the assumptions made.

Similar to macroscopic electrodynamics, the problem of construction of
models of gravitating media in macroscopic gravity is to be considered \cite
{Zala:Prog} after finding the macroscopic space-time geometry (see Section 
\ref{geometry}) and the form of averaged Einstein's 
operator (\ref{eq:averEE}). 

On using the tensor $U^{\alpha \beta }$ one can write the macroscopic field
equations in a remarkable form

\begin{equation}
\label{eq:aver-Einstein} G^{\alpha \epsilon }M_{\epsilon \beta }-\frac
12\delta _\beta ^\alpha G^{\mu \nu }M_{\mu \nu }=-\kappa T_\beta ^{\alpha 
{\rm (macro)}} 
\end{equation}
where the {\em macroscopic} stress-energy tensor $T_\beta ^{\alpha {\rm 
(macro)}}$ is defined to be 
\begin{equation}
\label{eq:macro-energy} \kappa T_\beta ^{\alpha {\rm (macro)}}=\kappa
 \langle {\bf t}_\beta ^{\alpha {\rm (micro)}}\rangle -(Z^\alpha {}_{\mu \nu
\beta }-\frac 12\delta _\beta ^\alpha Q_{\mu \nu })\bar g^{\mu \nu
}+U^{\alpha \epsilon }M_{\epsilon \beta }-\frac 12\delta _\beta ^\alpha
U^{\mu \nu }M_{\mu \nu }\ , 
\end{equation}
and the stress-energy tensor can be shown to satisfy the conservation law 
\begin{equation}
\label{eq:conservation} T_{\beta \parallel \alpha }^{\alpha {\rm (macro)}
}=0\ . 
\end{equation}
Though somewhat unexpected, the result seems to be natural: a space-time
averaging out of the Einstein equations brings the field equations which can
be written{\it \ in the form }of the Einstein equations for the induction
Ricci tensor defined through the Riemannian macroscopic metric $G_{\alpha
\beta }$. The macroscopic stress-energy tensor (\ref{eq:macro-energy})
includes, in addition to the averaged matter, the correlation tensor terms
with $Z^\alpha {}_{\mu \nu \beta }$, $Q_{\mu \nu }$ and $U^{\alpha \epsilon
} $ for geometric correction of the averaged matter. The correlation terms
reveal the structure of the correlation tensor $C_\beta ^\alpha $ in (\ref
{averEE:2}), but now all the correlation objects have geometrical origin and
definitions with corresponding differential equations for them (see (\ref
{eq:strZ}) and (\ref{eq:U})). In their geometrical meaning, the equations 
(\ref{eq:aver-Einstein}) however {\it are not Riemannian}, as the Einstein
equations (\ref{EE}) of general relativity are, which is due to 
a different underlying geometry for macroscopic gravitation.

The macrovacuum equations of the theory, following from the averaging of 
the vacuum Einstein equations $r_{\alpha \beta }=0$, or from (\ref{eq:averEE})
directly, read 
\begin{equation}
\label{eq:mvR:-Q}M_{\alpha \beta }=-Q_{\alpha \beta }\quad ,\quad Z^\alpha
{}_{\mu \nu \beta }\bar g^{\mu \nu }=-\bar g^{\alpha \epsilon }Q_{\beta
\epsilon }\ ,
\end{equation}
with $\bar g^{\alpha \beta }Q_{\alpha \beta }=0$ as a consequence. They show
the {\it Ricci non-flat} character of macroscopic gravitation in the
absence of averaged matter, in contrast to vacuum microscopic 
general relativity.   

\section{The correspondence principle and stress-energy of macroscopic
gravitational field}

The correspondence principle for the theory of macroscopic gravity \cite
{Zala:1993},\cite{Zala:1994} states that the macrovacuum equations (\ref
{eq:mvR:-Q}) become Isaacson's equations \cite{Isaa:1968} 
\begin{equation}
\label{eq:Isaa}M_{\alpha \beta }=-\kappa T_{\alpha \beta }^{{\rm (GW)}}
\end{equation}
in the high-frequency limit. As a result, the correlation tensor $Q_{\mu \nu
}=-2Z^\delta {}_{\mu [\nu }{}^\epsilon {}_{\underline{\delta }\epsilon ]}$,
which serves as the macrovacuum source in (\ref{eq:mvR:-Q}), is equal to $
\kappa T_{\alpha \beta }^{{\rm (GW)}}$ in the high-frequency limit and it
therefore describes the stress-energy of the macrovacuum gravitational
field. This result gives evidence in favour of considering the correlation
term on the right-hand side of the macroscopic field equations (\ref
{eq:averEE}) as the stress-energy tensor of macroscopic gravitation 
\begin{equation}
\label{eq:grav-energy}(Z^\alpha {}_{\mu \nu \beta }-\frac 12\delta _\beta
^\alpha Q_{\mu \nu })\bar g^{\mu \nu }=-\kappa T_\beta ^{\alpha {\rm (grav)}
}\ ,
\end{equation}
(which is $Z^\alpha {}_{\mu \nu \beta }\bar g^{\mu \nu }=-\kappa T_\beta
^{\alpha {\rm (grav)}}$ in the macrovacuum case (\ref{eq:mvR:-Q})). Indeed,
a simple consideration shows that averaging out over a space-time region
makes the gravitational field stress-energy localizable, which brings about
the corresponding {\em tensor} object. This fundamental fact has first been
established by Isaacson \cite{Isaa:1968} within the high frequency
approximation for general relativity. The macroscopic gravity approach
provides a general solution to the problem. The correlation tensor $Z^\alpha
{}_{\mu \nu \beta }$ (and $Q{}_{\mu \nu }=Z^\alpha {}_{\mu \nu \alpha }$ as a
consequence of algebraic properties of $Z^\alpha {}_{\beta [\gamma }{}^\mu
{}_{\underline{\nu }\lambda ]}$ \cite{Zala:1992},\cite{Zala:1993}) is just
an {\em averaged} effect of the ``product of connection'' which is known in
general relativity to be the stress-energy of gravitational field in its
physical meaning. Then the conservation law (\ref{eq:motion}) tells us that
only the {\em total} stress-energy of the averaged matter and the macroscopic
gravitational field is conserved
\begin{equation}
\label{eq:eq-motion}\left( \langle {\bf t}_\beta ^{\alpha {\rm (micro)}
}\rangle +T_\beta ^{\alpha {\rm (grav)}}\right) _{\parallel \alpha }=0\ .
\end{equation}
It should be pointed out here that there is a remarkable similarity of the
structure of the definition of $T_\beta ^{\alpha {\rm (grav)}}$ (\ref
{eq:grav-energy}) to the structure of Einstein's equations since $Z^\alpha
{}_{\mu \nu \beta }$ is a 
Ricci-tensor like object for the correlation tensor $
Z^\alpha {}_{\beta [\gamma }{}^\mu {}_{\underline{\nu }\lambda ]}$ and $
Q{}_{\mu \nu }=Z^\alpha {}_{\mu \nu \alpha }$ is an analogue of the curvature
scalar as a trace of $Z^\alpha {}_{\mu \nu \beta }$ (or, it is just the
Ricci tensor of $Q^\alpha {}_{\beta \gamma \delta }$). 

An analysis of the structure equations (\ref{eq:strZ}) for $Z^\alpha
{}_{\beta [\gamma }{}^\mu {}_{\underline{\nu }\lambda ]}$ shows that in this
simplest case the symmetric part $Z^\alpha {}_{(\mu \nu )\beta }$ remains
undetermined from the equations and 
\begin{equation}
\label{eq:strZ:conserv}Z^\alpha {}_{(\mu \nu )\beta \parallel \alpha }-\frac
12Q_{\mu \nu \parallel \beta }=0\ ,
\end{equation}
which means that the stress-energy tensor $T_\beta ^{\alpha {\rm (grav)}}$ is
conserved separately. Then, due to (\ref{eq:eq-motion}), or (\ref{eq:motion}),
the averaged matter stress-energy tensor 
$\langle {\bf t}_\beta ^{\alpha {\rm (micro)}}\rangle $ is 
also conserved separately . The fact of $
Z^\alpha {}_{(\mu \nu )\beta }$ being undetermined from (\ref{eq:strZ}) is
analogous to the fact that in the Riemannian geometry the Bianchi identities
do not determine the Ricci tensor which has to be fixed by an additional
hypothesis. In general relativity it is the Einstein equations which
relate the Ricci tensor part of the curvature to a matter distribution. Thus,
with one non-vanishing correlation tensor $Z^\alpha {}_{\beta [\gamma
}{}^\mu {}_{\underline{\nu }\lambda ]}$ the relations (\ref{eq:grav-energy})
must be taken as {\em the field equations} for the 
tensor $Z^\alpha {}_{(\mu \nu )\beta }$ with a {\em given}
stress-energy tensor $T_\beta ^{\alpha {\rm (grav)}}$, the
equations being algebraic. 

When the higher correlation tensors, $Y^\alpha {}_{\beta [\gamma }{}^\mu {}_{
\underline{\nu }\sigma }{}^\theta {}_{\underline{\kappa }\pi ]}$ and $
X^\alpha {}_{\beta [\gamma }{}^\mu {}_{\underline{\nu }\sigma }{}^\theta {}_{
\underline{\kappa }\pi }{}^\tau {}_{\underline{\varphi }\psi ]}$, are taken
into account a distribution of the stress-energy tensor $T_\beta ^{\alpha 
{\rm (grav)}}$ is to be calculated from (\ref{eq:grav-energy}) by solving 
the generalized analogue of
equations (\ref{eq:strZ}) for $Z^\alpha
{}_{\beta [\gamma }{}^\mu {}_{\underline{\nu }\lambda ]}$ 
\cite{Zala:1992},\cite{Zala:1993}, and the
conservation law (\ref{eq:eq-motion}) holds. In such a case one should find
some additional hypotheses for the higher 
correlation tensors \cite{Zala:Prog}. 

\section{Using Einstein's equations in cosmology}

A study of the structure of the field equations of macroscopic gravity
enables one to answer also a fundamental question of cosmology about the
physical meaning and the range of applicability of the 
Einstein equations with a
continuous (smoothed) matter source. Indeed, if {\it all correlations
functions vanish, }$Z^\alpha {}_{\beta [\gamma }{}^\mu {}_{\underline{\nu }
\sigma ]}=0$ (connection correlations) and $U^{\alpha \beta }=0$ (metric
correlations) the equations (\ref{eq:aver-Einstein}) (as well as (\ref
{eq:mvR:-Q})) become the Einstein equations 
\begin{equation}
\label{eq:aver-Einstein:cosm} G^{\alpha \epsilon }M_{\epsilon \beta }-\frac
12\delta _\beta ^\alpha G^{\mu \nu }M_{\mu \nu }=-\kappa T_\beta ^{\alpha 
{\rm (macro)}} 
\end{equation}
for {\it the macroscopic metric} $G_{\alpha \beta }$ with a stress-energy 
tensor 
\begin{equation}
T_\beta ^{\alpha {\rm (macro)}}=\langle {\bf t}_\beta ^{\alpha {\rm (micro)}
}\rangle
\end{equation}
on the right-hand side of (\ref{eq:aver-Einstein:cosm}), which 
is usually taken as a perfect fluid
tensor while looking for cosmological solutions. 
This reveals that the physical
meaning and essence of using Einstein's equations in studies of cosmological
problems consists in neglecting the gravitational field correlations.

To take into account the gravitational field correlations and to 
understand their
effect and role in the dynamics of the universe, the equations of macroscopic
gravity are to be solved. One of the most important questions here is to
clear up the status of the homogeneous and isotropic
(Friedmann-Robertson-Walker) models in cosmology. Given a hydrodynamic
stress-energy tensor $\langle {\bf t}_\beta ^{\alpha ({\rm micro})}\rangle $ 
(or upon calculating it for 
a given microscopic gravitating matter model) and an
equation of state, there is the following system of partial differential
equations in the simplest case then only the correlation tensor 
$Z^\alpha {}_{\beta [\gamma }{}^\mu {}_{\underline{\nu }\sigma ]}$ is 
taken into account: the field
equations (\ref{eq:averEE}), or (\ref{eq:aver-Einstein}), for the
macroscopic metric tensor $G_{\alpha \beta }$, the equations (\ref{eq:strZ})
for $Z^\alpha {}_{\beta [\gamma }{}^\mu {}_{\underline{\nu }\sigma ]}$
together with the field equation (\ref{eq:grav-energy}) with a given, or
assumed, stress-energy tensor of macroscopic gravitational field $T_\beta
^{\alpha {\rm (grav)}}$, and the equations (\ref{eq:U}) for $U^{\alpha \beta
}$. 

This is the system for that part of macroscopic gravity which is 
related to the
Riemannian induction fields $G_{\alpha \beta }$, or 
$M^{\alpha}{}_{\beta \gamma \delta}$, 
and correlation fields $Z^\alpha {}_{\beta [\gamma }{}^\mu {}_{
\underline{\nu }\sigma ]}$ and $U^{\alpha \beta }$, the part being closed in
both geometry and field equations\footnote{The non-trivial
fact of decoupling of the induction and average field parts of the presented
formulation of macroscopic gravity is a consequence of two assumptions: (A)
the averaged microscopic tensor $\bar r^\alpha {}_{\beta \gamma \delta }$ is
supposed to be again a curvature tensor of a metric affine connection
geometry with curvature $R^\alpha {}_{\beta \gamma \delta }=\bar r^\alpha
{}_{\beta \gamma \delta }$ (see Section \ref{geometry}), which allows one to
apply Schouten's classification \cite{Scho-Stru:1935} of such geometries
where the metric and non-metric parts {\em do decouple}, and (B) the
correlations between metric and connection are assumed to vanish 
(\ref{aver:gF}),
which preserves in fact the decoupling mentioned in (A). It should be
emphasized here that the rule (\ref{aver:gF}) brings 
nevertheless the non-trivial
correlations between metric and curvature (\ref{eq:gr}). On the other
side, it is that rule that makes it possible to write down the averaged
Einstein equations (\ref{eq:averEE}) in the form of Einstein's equations 
(\ref{eq:aver-Einstein}). Such a structure of the theory of macroscopic
gravity is one of the simplest possible formulations, 
and any generalization either (A)
or (B) leads to more sophisticated macroscopic space-time geometries and
field equations, physical content and interpretation of which are much more
difficult \cite{Zala:Prog}.}. The second part of macroscopic 
gravity is non-Riemannian and
related to the average fields which are represented by the connection $\Pi
^\alpha {}_{\beta \gamma }$, or the curvature $R^\alpha {}_{\beta \gamma
\delta }$, the affine deformation tensor $A^\alpha {}_{\beta \gamma }$ and
the non-metricity object $N{}_{\alpha \beta \gamma }$. To find all these
objects after resolving the first part one should solve the equation (\ref
{eq:A}) for $A^\alpha {}_{\beta \gamma }$ with a determined correlation
tensor $Q^\alpha {}_{\beta \gamma \delta }$ with subsequent calculations of $
\Pi ^\alpha {}_{\beta \gamma }=\overline{{\cal F}}^\alpha {}_{\beta \gamma
}-A^\alpha {}_{\beta \gamma }$, $N{}_{\alpha \beta \gamma }$ and $R^\alpha
{}_{\beta \gamma \delta }$.

Due to the physical structure of macroscopic gravity as a classical
macroscopic theory with one-point averages it can be applied for the 
description of the universe since decoupling matter and radiation when
the evolving universe can be considered as being in an equilibrium state. 

\section*{Acknowledgments}

It is a pleasure for me to thank Dennis Sciama for hospitality and support
during my visits to SISSA and ICTP where the work
has been done in part, and Boris Dubrovin, George Ellis and Reza Tavakol for
many discussions and encouragement. I would like to express my gratitude to
Ian Roxburgh for his hospitality in the School of Mathematical Sciences, 
QMW. I am thankful to Jorge Devoto, Chris Isham, 
Nelly Konopleva, Andrzej Krasi\'nski, Roy Maartens, Malcolm MacCallum, 
David Matravers, Alexei Nesteruk, Kamilla Piotrkowska, and Bill Stoeger for 
discussions. I thank Henk van Elst for careful reading of the manuscript 
and discussion. The author was supported by a Royal Society fellowship.

\end{document}